\documentclass[%
preprint,
 amsmath,amssymb,
 aps,
]{revtex4-1}
\usepackage{graphicx}
\graphicspath{{./Figures/}}
\usepackage{slashed}
\usepackage{graphicx}
\usepackage{dcolumn}
\usepackage{bm}

\begin{document}


\title{Analysis of radial excitations of octet baryons in QCD sum rules}

\author{T.M.Aliev}
  \email{taliev@metu.edu.tr}
 \author{S. Bilmis}%
\affiliation{%
Department of Physics, Middle East Technical University, 06800, Ankara, Turkey\\
}%

\date{\today}

\begin{abstract}
Using the QCD sum rules method, we estimate the mass and residues of the first radial excitations of octet baryons. The contributions coming from the ground state baryons are eliminated by constructing the linear combinations of the sum rules corresponding to different Lorentz structures. Our predictions of the masses of the first radial excitations of octet baryons are in good agreement with the data.

\end{abstract}

\maketitle


\section{\label{sec:level1}Introduction}
At present time, many radial excitations of the mesons and baryons which carry the same spin-parity quantum numbers as the ground states are observed in experiments~\cite{pdg}.
The identification and investigation of the properties of the radial excitations on the background of the hadronic continuum  is a quite difficult task in the experiments. These resonances are strongly coupled to the two, three hadrons, which leads to the large decay widths. Theoretical study of the radial excitations is also a challenging problem. One of the powerful theoretical methods for studying the properties of the hadrons is the sum rules method. The QCD sum rules method introduced in~\cite{shifman1979qcd} was extremely useful for studying the properties of mesonic systems.
This method was extended to ground state baryons in pioneering work~\cite{Ioffe:1981kw}. It is well known that the choice of interpolating current which carries the same quantum numbers as appropriate baryon is not uniquely defined. The most general form of the interpolating currents for octet baryons is found in~\cite{chung1982baryon}. It is contemplated that whether this method can be applied to the radial excitations of baryons. Note that the radial excitation of mesons within finite energy rules have been studied in~\cite{krasnikov1982influence,gorishny1984next}. Recently, radial excitations of heavy-light mesons have been investigated in details via the QCD sum rules in~\cite{gelhausen2014radial}. Recently, this method  was applied for the determination of the masses of first radial excitation of mesons and nucleon by using the least square fitting method~\cite{jiang2015radial}.
In this paper, we calculate the mass and residues of the first excited states of octet baryons within the QCD sum rules method. We modify the spectral representation from the hadronic part by taking into account two poles which correspond to the ground and first excited state baryons, respectively. Then, the method for eliminating the ground state contributions is employed for the determination of the mass and residues of the excited state baryons.

The paper is organized as follows. In Section~\ref{sec:level2}, we derive the mass sum rules for octet baryons by including the first radial excitation baryons. Section~\ref{sec:level3} is devoted to the numerical analysis of obtained rules. Section~\ref{sec:conclusion} contains the summary of our results and conclusions.

\section{\label{sec:level2} Mass sum rules for octet baryons with their radial excitation}
For determination of the mass and residues of the mesons and baryons, usually two-point correlation function is used. Following the sum rules method strategy for extracting the mass and residues of radially excited octet baryons, we consider the following two-point correlation function
\begin{equation}
  \label{eq:corr}
  \Pi (q) = i \int d^4x~e^{ipx} \langle 0|T\{\eta(x) \bar{\eta}(0)\}|0 \rangle~,
\end{equation}
where $\eta(x)$ is the interpolating currents of the octet baryons. The most general form of the interpolating currents for the octet baryons are~\cite{chung1982baryon}:
\begin{equation}
  \label{eq:currents}
  \begin{split}
    \eta_p(x) &= 2 \epsilon^{abc} \sum_{l=1}^2 \big((u^a)^T C A_1^l d^b \big) A_2^l u^c~, \\
    \eta_n(x) &= \eta_p (u \leftrightarrow d)~, \\
    \eta_{\Sigma^+}(x) &= \eta_p (d \rightarrow s)~, \\
    \eta_{\Sigma^-}(x) &= \eta_n (u \rightarrow s)~, \\
    \eta_{\Xi^0}(x) &= \eta_{\Sigma^+} (u \leftrightarrow s)~, \\
    \eta_{\Xi^-}(x) &= \eta_{\Sigma^-} (d \leftrightarrow s)~, \\
    \eta_{\Sigma^0}(x) &= \sqrt{2} \epsilon^{abc} \sum_{l=1}^{2} \big[ (u^{aT} C A_1^l s^b) A_2^l d^c + (d^{aT} C A_1^l s^b) A_2^l u^c \big]~,
  \end{split}
\end{equation}
where $A_1^1 = I$,~$A_1^2 = A_2^1 = \gamma^5$,~$A_2^2 = \beta$ in which $\beta$ is arbitrary parameter and $a,b$ and $c$ are color indices.

The phenomenological part of the correlation function can be obtained by inserting a full set of baryons carrying the same quantum numbers as the interpolating current. Isolating the ground state and its first radial excitation from phenomenological side for the correlation functions we have
\begin{equation}
  \label{eq:piphys}
  \Pi^{phys} = \frac{\lambda^2 (\slashed{p} + m)}{p^2 - m^2} + \frac{\lambda_1^2 (\slashed{p} + m)}{p^2 - m_1^2} + ...
\end{equation}

Here $\lambda (\lambda_1)$ and $m (m_1)$ are the residue and mass of the ground (first radial excitation) state baryon and $...$ stands for the contributions of higher states and continuum. In derivation of eq.~\eqref{eq:piphys}, we used
\begin{equation}
  \label{eq:B}
  \langle  0 | \eta | B(p) \rangle  \equiv \lambda u(p)~.
\end{equation}

In order to suppress the higher states and continuum contributions the Borel transformation is applied. After performing Borel transformation from eq.~\eqref{eq:piphys} we have,

\begin{equation}
  \label{eq:Bpsqr}
  B_{p^2} \Pi^{phys} \equiv \lambda_p^2 (\slashed{p} +m) e^{-m^2/M^2} + \lambda_1^2 (\slashed{p} + m_1) e^{-m_1^2/M^2} + ...
\end{equation}

The correlation function in terms of quark-gluon degrees of freedom  (theoretical part) is calculated in the deep Euclidean domain ($p^2 << 0$) with the help of the operator product expansion (OPE), which contains the perturbative and non-perturbative (vacuum condensate) contributions. For structures $\slashed{p}$ and $I$ it can be written as;
\begin{equation}
  \label{eq:ope}
  \Pi_i^{OPE}(p^2) = \Pi_i^{(pert)}(p^2) + \Pi_i^{(non-pert)}(p^2) 
\end{equation}
where the invariant functions $\Pi_1~(\Pi_2)$ correspond to the coefficient of the structure $\slashed{p}~(I)$. 

The expressions of various terms entering to the right side of the eq.~\eqref{eq:ope} are calculated in numerous works (see for example ~\cite{lee2002predictive,aliev2002octet}). Performing Borel transformation over $p^2$ in theoretical part of the correlation function and equating the coefficients of the Lorentz structures $\slashed{p}$ and $I$, one can obtain the sum rules for the mass.

The sum rules for the structures $\slashed{p}$ and $I$ can be written as
\begin{equation}
  \label{eq:pstructure}
  \begin{split}
    \Pi_1(M^2) &= \lambda^2 e^{-m^2/M^2} + \lambda^{\prime 2} e^{- m^\prime / M^2} \\
    & = C_1 E_2 M^6 + C_2 m_s \langle \bar{q}q\rangle  E_0 M^2 + C_3 m_s m_0^2 \langle \bar{q}q\rangle  \\
    &+ C_4 \langle qq\rangle ^2 + C_5 m_0^2 \langle  q q\rangle ^2/M^2 + C_6 \langle \frac{\alpha_S}{\pi} G^2\rangle  M^2 ,\\
    \\
        \Pi_2(M^2) & = \lambda^2 m e^{-m^2/M^2} + \lambda^{\prime 2}m^\prime e^{-m^{\prime 2}/M^2} \\
    & C_1^\prime m_s M^6 E_2(x) + C_2^\prime \langle \bar{q}q\rangle  M^4 E_1 + C_3^\prime m_0^2 \langle \bar{q}q\rangle  E_0 M^2 \\
    &+ C_4^\prime m_s \langle qq\rangle ^2 + C_5^ \prime m_s \langle  \frac{\alpha_s}{\pi} G^2 \rangle  M^2 + C_6^\prime \langle \bar{q}q\rangle  \langle \frac{\alpha_s}{\pi} G^2\rangle ~.
\end{split}
\end{equation}
In eq.~\eqref{eq:pstructure}, we consider the contributions of operators up to dimension $6$. Obviously, $C_i$ for different members of the octet baryons are different. Note that in the derivation of eq.~\eqref{eq:pstructure}, we set the light quark masses to zero, $m_u= m_d=0$.
The continuum subtraction is done using the quark-hadron duality ansatz. In eq.~\eqref{eq:pstructure} the function $E_2 (s_0/M^2)$ is described of the by higher states contributions and continuum contributions and determined as $E_2 (x) = 1 - e^{-x} \sum{\frac{x^n}{n!}}$ where $s_0$ is the effective continuum threshold. The coefficients $C_i$ in eq.~\eqref{eq:pstructure} are given by (see~\cite{lee2002predictive,aliev2002octet}):
\\ For $N$
\begin{equation}
  \label{eq:c}
  \begin{split}
    C_1 &= \frac{1}{256 \pi^4} (5 + 2\beta + 5\beta^2)~, \\
    C_2 &= 0~,~~C_3=0~, \\
    C_4 &= \frac{1}{6} \big[ 6(-1 + \beta^2) + (-1+\beta)^2 \big]~, \\
    C_5 &= -\frac{1}{24} \big[ 12 (-1 + \beta^2) + (-1+\beta)^2 \big]~, \\
    C_6 &= \frac{1}{256 \pi^2} (5 \beta^2 + 2\beta +5)~.
  \end{split}
\end{equation}
For $\Sigma$
\begin{equation}
  \label{eq:csigma}
  \begin{split}
    C_1 &= \frac{1}{256 \pi^4} (5 + 2\beta + 5\beta^2)~, \\
    C_2 &= \frac{1}{32\pi^2} \big[ (5 + 2\beta + 5 \beta^2) \gamma -12(-1+\beta^2) \big]~, \\
    C_3 &= \frac{1}{96 \pi ^2} \big(\gamma (4+4\beta+ 4\beta^2) + 21(\beta^2 -1) \big) + \frac{3}{16 \pi^2}(-1+\beta^2) \big(\gamma_E - \ln \frac{M^2}{\mu^2} \big)~, \\
    C_4 &= \frac{1}{6} \big[ 6(-1 + \beta^2) \gamma + (-1+\beta)^2 \big]~, \\
    C_5 &= -\frac{1}{24} \big[ 12 (-1 + \beta^2) \gamma + (-1+\beta)^2 \big]~, \\
    C_6 &= \frac{1}{256 \pi^2} (5 \beta^2 + 2\beta +5)~.
  \end{split}
\end{equation}
For $\Xi$;
\begin{equation}
  \label{eq:cXi}
  \begin{split}
    C_1 &= \frac{1}{256 \pi^4} (5 + 2\beta + 5\beta^2)~, \\
    C_2 &= \frac{3}{16 \pi^2} \big( -2(-1 + \beta^2) +(1+\beta)^2\gamma \big)~, \\
    C_3 &= \frac{1}{96 \pi ^2} \big( (15-\gamma)\beta^2 - 10 \gamma \beta - (15+f)\big) + \frac{3}{16 \pi^2} (-1 + \beta^2) \big(\gamma_E - \ln \frac{M^2}{\mu^2} \big)~, \\
    C_4 &= \frac{f}{6} \big[ 6(-1 + \beta^2) + (-1+\beta)^2 \gamma \big]~, \\
    C_5 &= -\frac{1}{24} \gamma \big[ 12 (-1 + \beta^2) + (-1+\beta)^2 \gamma \big]~, \\
    C_6 &= \frac{1}{256 \pi^2} (5 \beta^2 + 2\beta +5)~.
  \end{split}
\end{equation}
For $\Lambda$
\begin{equation}
  \label{eq:cLambda}
  \begin{split}
    C_1 &= \frac{1}{256 \pi^4} (5 + 2\beta + 5\beta^2)~, \\
    C_2 &= \frac{1}{96 \pi^2} \big( 3(5 + 2\beta + 5\beta^2)\gamma + 4(1 + 4\beta - 5\beta^2) \big)~, \\
    C_3 &= -\frac{1}{96 \pi ^2} \big( 4( 1 + \beta + \beta^2)\gamma + (-5\beta^2 + 4\beta +1) \big)  - \frac{1}{16 \pi^2} (1-\beta^2) \big(\gamma_E - \ln \frac{M^2}{\mu^2} \big)~, \\
    C_4 &= -\frac{1}{18} (1-\beta) \big( 2\gamma (1+5\beta) + (13 + 11\beta) \big)~, \\
    C_5 &= \frac{1}{72}(1-\beta) \big( 8(1+2\beta)\gamma + (25+23\beta) \big)~, \\
    C_6 &= \frac{1}{256 \pi^2} (5 \beta^2 + 2\beta +5)~.
  \end{split}
\end{equation}
Moreover, the coefficients $C_i^\prime$ for the members of the octet baryons are given as follows.
\\
For $N$
\begin{equation}
  \label{eq:cnprime}
  \begin{split}
    C_1^\prime &= 0~, \\
    C_2^\prime &= -\frac{1}{16 \pi^2} ( 7 \beta^2 - 2\beta -5 )~, \\
    C_3^\prime &= \frac{3}{16 \pi ^2} (\beta^2 -1)~, \\
    C_4^\prime &= 0~, \\
    C_5^\prime &= 0~, \\
    C_6^\prime &= -\frac{1}{288} (19 \beta^2 + 10\beta -29)~.
  \end{split}
\end{equation}
For $\Sigma$
\begin{equation}
  \label{eq:csigmaprime}
  \begin{split}
    C_1^\prime &= \frac{1}{64 \pi^4} (\beta - 1)^2~, \\
    C_2^\prime &= -\frac{1}{16 \pi^2} \big( (6+\gamma) \beta^2 -2\gamma \beta -(6-f) \big)~, \\
    C_3^\prime &= \frac{3}{16 \pi ^2} (\beta^2 -1)~, \\
    C_4^\prime &= \frac{1}{6} \big( (5-3\gamma) \beta^2  +2\beta + (5+3\gamma) \big)~, \\
    C_5^\prime &= - \frac{1}{128 \pi^2} (1-\beta)^2~, \\
    C_6^\prime &= -\frac{1}{288} \big( (24-5\gamma) \beta^2 + 10 \gamma \beta - (24 + 5\gamma) \big)~.
  \end{split}
\end{equation}
For $\Xi$
\begin{equation}
  \label{eq:cxiprime}
  \begin{split}
    C_1^\prime &= \frac{3}{32 \pi^4} (\beta^2 - 1)~, \\
    C_2^\prime &= -\frac{1}{16 \pi^2} \big( (6\gamma+1) \beta^2 -2\beta -(6\gamma-1) \big)~, \\
    C_3^\prime &= \frac{3\gamma}{16 \pi ^2} (\beta^2 -1)~, \\
    C_4^\prime &= \frac{\gamma}{2} \big( (3-\gamma) \beta^2 + 2\beta + 3 + \gamma \big)~, \\
    C_5^\prime &= \frac{3}{64 \pi^2} (\beta^2-1)~, \\
    C_6^\prime &= -\frac{1}{288} \big( (24 \gamma-5) \beta^2 + 10 \beta - (24\gamma + 5) \big)~.
  \end{split}
\end{equation}
For $\Lambda$
\begin{equation}
  \label{eq:cLambdaprime}
  \begin{split}
    C_1^\prime &= \frac{1}{192 \pi^4} (11\beta^2 +2\beta -13)~, \\
    C_2^\prime &= -\frac{1}{48 \pi^2} \big( (10+11\gamma) \beta^2 +(-8+2\gamma)\beta - (2+13\gamma) \big)~, \\
    C_3^\prime &= -\frac{1}{16 \pi ^2} \big( (-1-2\gamma)\beta^2 + (1+2\gamma) \big)~, \\
    C_4^\prime &= \frac{1}{18} \big( (15-5 \gamma) \beta^2 + (6+4\gamma) \beta + (15 + \gamma) \big)~,\\
    C_5^\prime &= \frac{1}{384 \pi^2} (13 \beta^2- 2\beta -11)~, \\
    C_6^\prime &= -\frac{1}{864} \big( ( 4 + 53\gamma) \beta^2 + (40 -10 \gamma) \beta - (44 + 4 \gamma) \big)~.
  \end{split}
\end{equation}
where $\gamma_E$ is the Euler constant, $\gamma = \frac{\langle \bar{s}s\rangle}{\langle \bar{u}{u} \rangle}$ and $\mu$ is the renormalization scale parameter whose value is taken as $\mu = 0.5~\rm{GeV}$. Now using
eqs.~\ref{eq:Bpsqr} and \ref{eq:pstructure} one can determine the masses and residues of the octet baryons. We have four unknowns (two masses and two residues). Therefore we need four equations to find these unknowns.

First two equations are obtained by equating the coefficients of the structures $\slashed{p}$ and $I$ in both representations of the correlation functions i.e.

\begin{equation}
  \label{eq:pionepitwo}
  \begin{split}
  \lambda^2 e^{-m^2/M^2} + \lambda^{\prime 2} e^{-m^{\prime 2}/M^2} &= \Pi_1~, \\ 
  \lambda^2 m e^{-m^2/M^2} + \lambda^{\prime 2} m^\prime e^{-m^{\prime 2}/M^2} &= \Pi_2~.  
\end{split}
\end{equation}

The remaining two equations can be obtained by taking derivatives with respect to $-1/M^2$ from both sides of the eq.~\eqref{eq:pionepitwo}. Then we have,

\begin{equation}
  \label{eq:pionepitwoderivative}
  \begin{split}
  m^2 \lambda^2 e^{-m^2/M^2} + \lambda^{\prime 2} m^{\prime 2} e^{-m^2/M^2} &= \Pi_{1}^\prime~, \\ 
  \lambda^2 m^3 e^{-m^2/M^2} + \lambda^{\prime 2} m^{\prime 3} e^{-m^2/M^2} &= \Pi_2^\prime~.  
\end{split}
\end{equation}

From these equations we get;

\begin{equation}
  \label{eq:primesquare}
  \begin{split}
    m^{\prime 2} &= \frac{\Pi_2^\prime - m \Pi_{1}^\prime} {\Pi_2 - m \Pi_{1}}~, \\
    \lambda^{\prime 2} &= \frac{1}{m^{\prime 2} - m^2} e^{m^{\prime 2}/M^2} [\Pi_{1}^{\prime} - m^2 \Pi_1]
  \end{split}
\end{equation}

To obtain the mass and residues of the radial excitations from the sum rules, we take the mass of the ground state as an input parameter. Note that the zero width approximation is assumed.

\section{\label{sec:level3}Numerical Analysis}
The input parameters used in the above coefficients include the strange quark mass and quark condensates. In our numerical calculations, we use following values of these parameters see~\cite{Narison:2005zg,Ioffe:1981kw,Narison:2009vy}:

\begin{equation}
  \label{eq:parameters}
  \begin{split}
    m_0^2 &= (0.8 \pm 0.2)~\rm{GeV^2}, \\
    \langle \bar{q} q\rangle  (2~\rm{GeV}) &= - (277^{+12}_{-10}~\rm{MeV})^3~, \\
    m_s(2~\rm{GeV}) &= (95 \pm 5)~\rm{MeV}~, \\
    \gamma = \frac{\langle \bar{s}s\rangle }{\langle \bar{q}q\rangle } &= (0.8 \pm 0.2)~, \\
   \kappa \alpha_s \langle \bar{q}q\rangle^2 &= (5.8 \pm 1.8) \times 10^{-4}~\rm{GeV}^6~~.
  \end{split}
\end{equation}
In numerical calculations we take $\kappa=1$.

The sum rules for the mass and residues of the radially excited baryons contain three auxiliary parameters in addition to these input parameters; the Borel mass $M^2$, arbitrary parameter $\beta$ (in expressions of the interpolating current) and continuum threshold $s_0$. Usually the continuum threshold is related to the energy of the first excited state. In our calculations for $s_0$ we have used $\sqrt{s_0} = m_{ground} + \Delta$~\rm{GeV} where $\Delta$ varies between $0.3 - 0.8$. These values of $s_0$ include only the mass of the first radial excitation, while the higher excitations are included to the continuum states. The working interval of $M^2$ is obtained by the following way. The lower bound of $M^2$ is determined by demanding the convergence of the operator product expansion, while the upper bound is obtained by requiring the continuum contribution remains subleading. Our calculations lead to the following working region of $M^2$:
\begin{equation}
  \label{eq:region}
  \begin{split}
    N&:\hspace{2mm} 1.2~\rm{GeV}^2  \leq M^2  \leq 2.2~\rm{GeV}^2~, \\
    \Sigma&:\hspace{2mm} 1.4~\rm{GeV}^2  \leq M^2 \leq 2.4~\rm{GeV}^2~, \\
     \Xi&:\hspace{2mm} 1.8~\rm{GeV}^2  \leq M^2 \leq 3~\rm{GeV}^2~, \\
     \Lambda&:\hspace{2mm} 1.4~\rm{GeV}^2  \leq M^2 \leq 2.4~\rm{GeV}^2~. 
  \end{split}
\end{equation}

\begin{figure}[h]
  \includegraphics[scale=0.7]{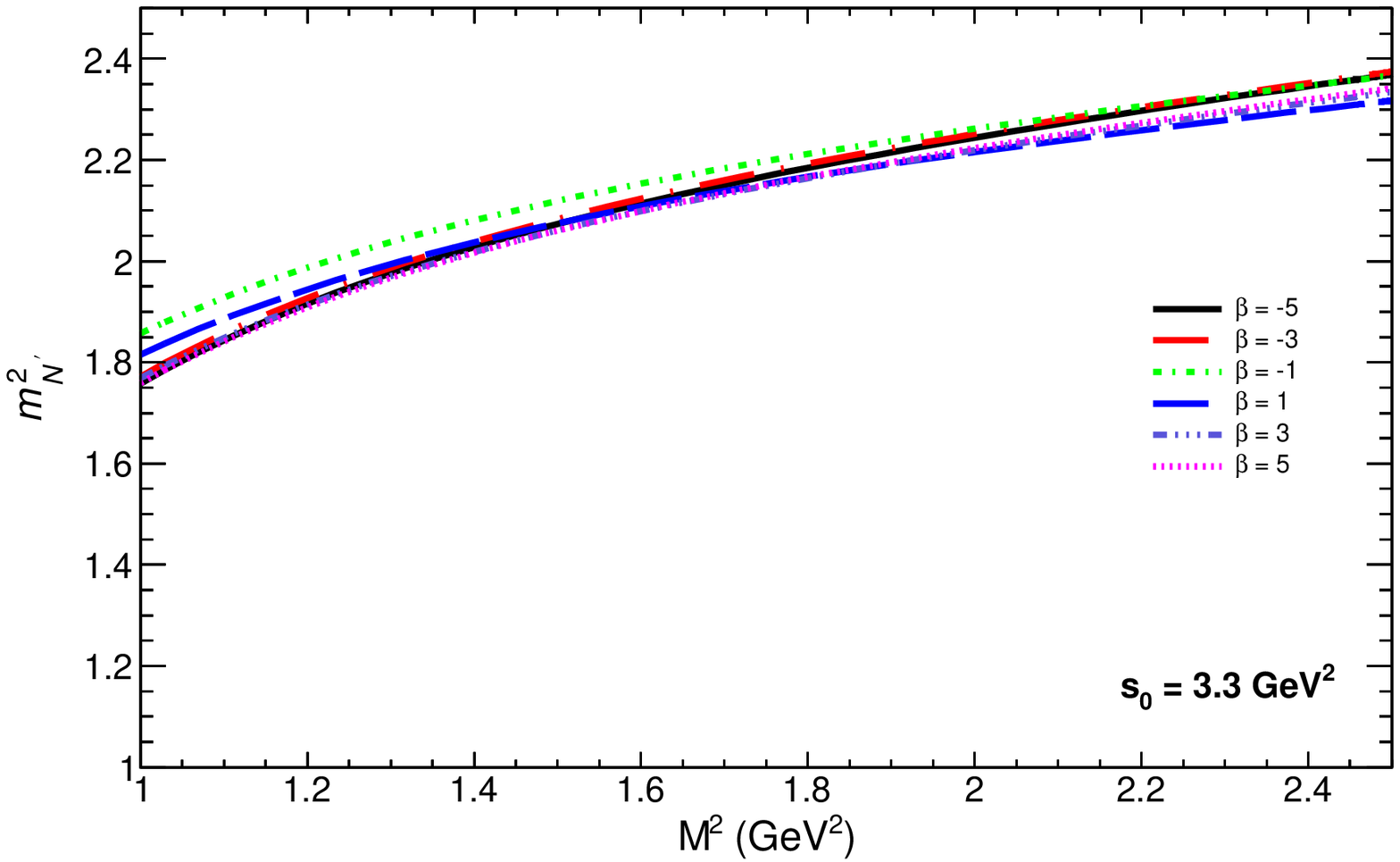}
  \caption{Dependence of the $m_{N^\prime}^2$ on the Borel mass parameter $M^2$ for the fixed value of the continuum threshold $s_0=3.3~\rm{GeV^2}$ is depicted for the several fixed values of parameter $\beta$.}
  \label{fig:one}
\end{figure}

\begin{figure}[h]
  \includegraphics[scale=0.7]{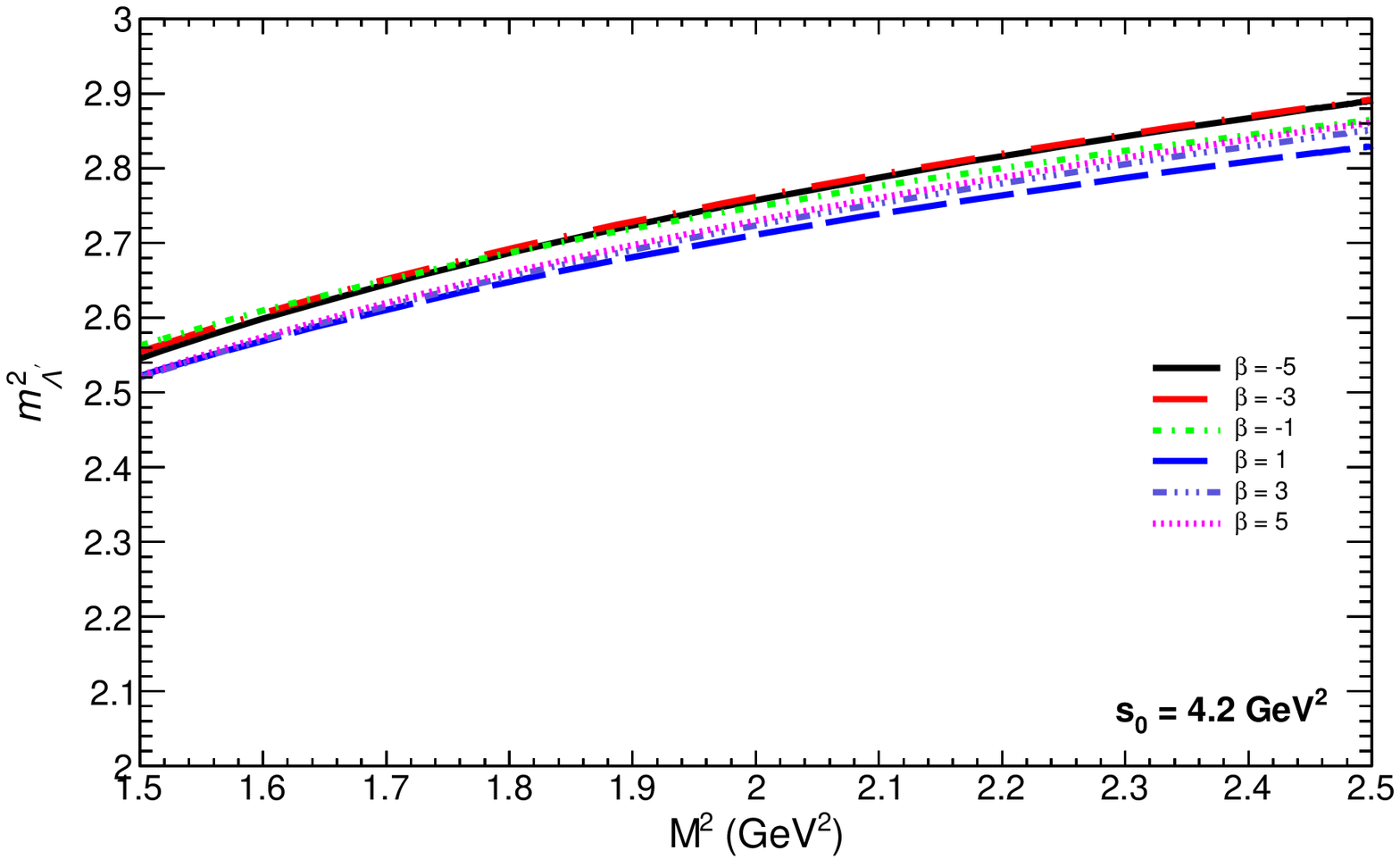}  
  \caption{Dependence of the $m_{\Lambda^\prime}^2$ on the Borel mass parameter $M^2$ for the fixed value of the continuum threshold $s_0=4.2~\rm{GeV^2}$ is depicted for the several fixed values of parameter $\beta$.}
  \label{fig:two}
\end{figure}

Having determined the working regions for Borel mass parameter $M^2$, we can calculate the mass and residues of the first radial excitation of octet baryons. As an example, in Figures~\ref{fig:one} and \ref{fig:two}, we present the dependence of the mass of the radial excitation of $N$ and $\Lambda$ baryons on $M^2$ at fixed values of $\beta$ and $s_0$. From these figures, we obtained that the masses of the radial excitations of $N$ and $\Lambda$ exhibit good stability with respect to the variation of $M^2$ in the working region. We perform the same analysis for another fixed values of $s_0$ and find that the results change about $3.5~\%$. We also performed analysis for the other members of the octet baryons and obtained that the dependence of the mass of radial excitation of octet baryons on $M^2$ is rather weak.

In order to determine the optimal working interval of the parameter $\beta$, we study the dependence of the mass and residues of octet baryons and their radial excitation on $\cos{\theta}$, where $\beta = \tan{\theta}$. Note that we use $\cos{\theta}$ by the following reason. Exploring the whole region in $\beta$ ($-\infty,\infty$) is equivalent to the very restricted domain $(-1,1)$.

\begin{figure}[h]
  \includegraphics[scale=0.7]{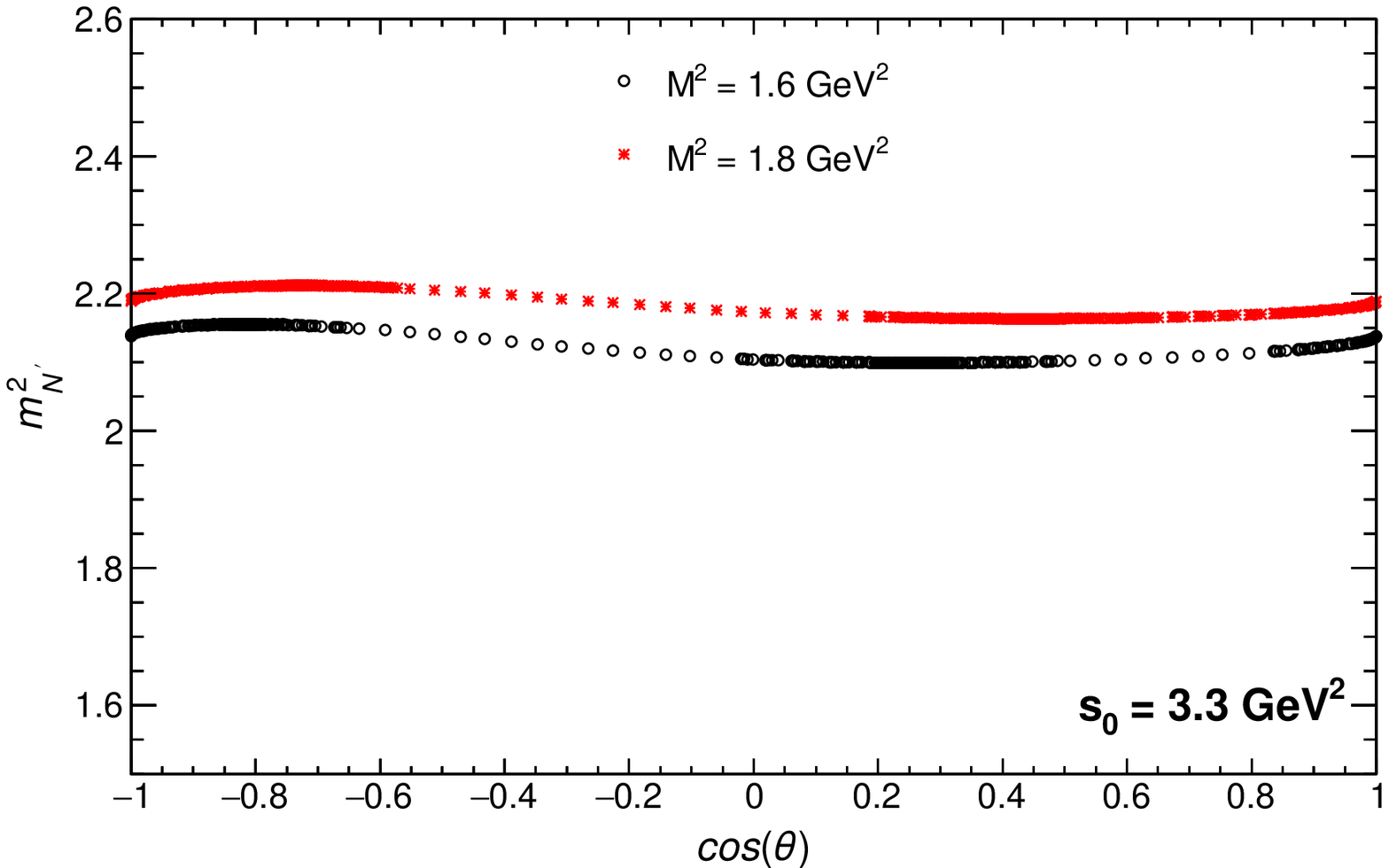}
  \caption{Dependence of the $m_{N^\prime}^2$ on $\cos{\theta}$ at the fixed value of the continuum threshold $s_0 = 3.3~\rm{GeV^2}$ is shown for various values of the Borel mass parameter $M^2$.}
\label{fig:three}
\end{figure}

\begin{figure}[h]
  \includegraphics[scale=0.7]{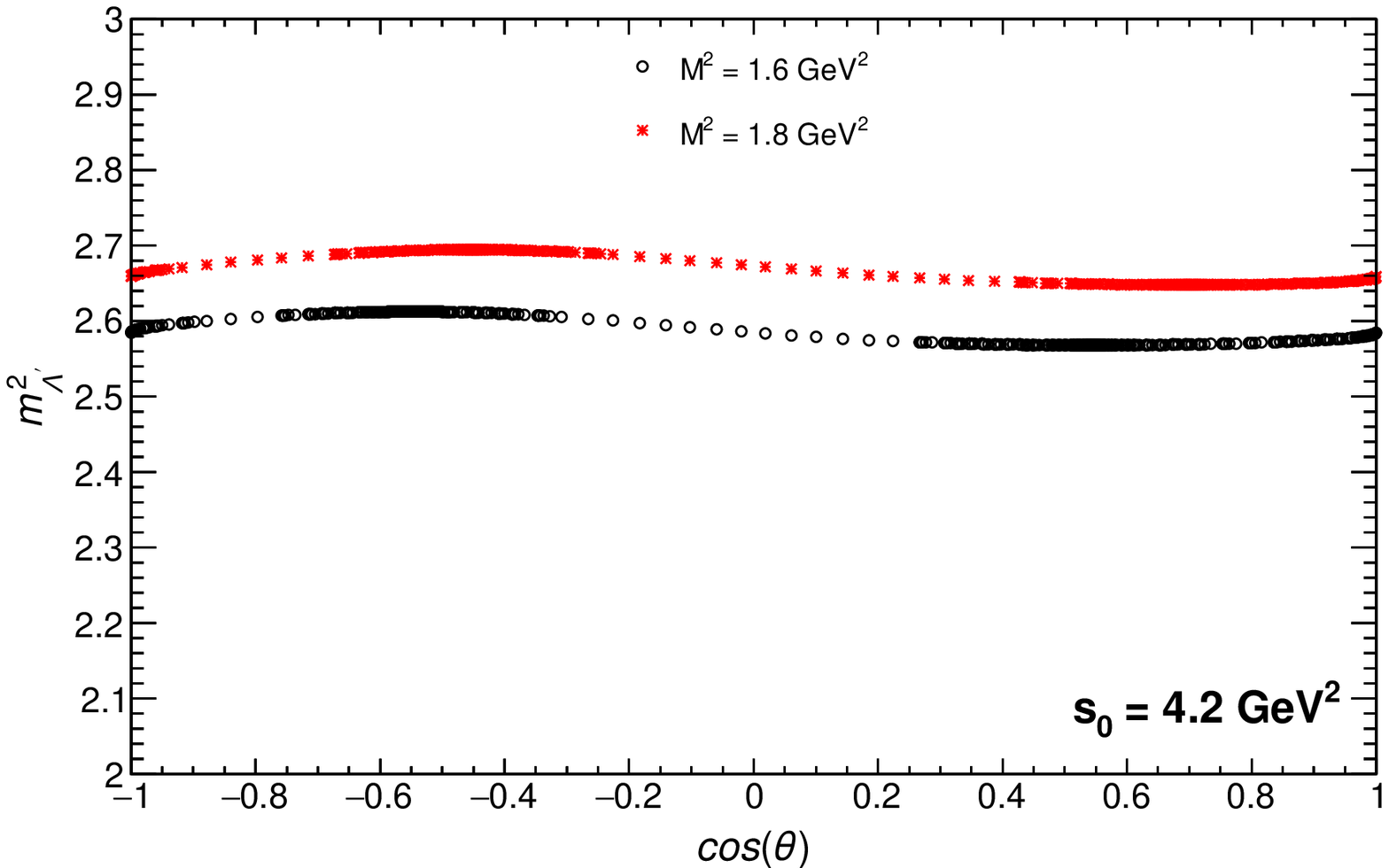}
  \caption{Dependence of the $m_{\Lambda^\prime}^2$ on $\cos{\theta}$ at the fixed value of the continuum threshold $s_0 = 4.2~\rm{GeV^2}$ is shown for various values of the Borel mass parameter $M^2$.}
\label{fig:four}
\end{figure}

In Figures~\ref{fig:three} and \ref{fig:four}, we present the dependence $m_{N^\prime}^2$ and $m_{\Lambda^\prime}^2$ on $\cos {\theta}$ for two fixed values of $M^2$ and at fixed value $s_0$. From these figures, we observe that in the domain $-0.6 \leq \cos \theta \leq 0.8$ the mass of $N^\prime$ and $\Lambda^\prime$ baryons remains stable with respect to the variation of $\cos \theta$ and we deduce following values for their masses;

\begin{equation}
  \label{eq:massprime}
  \begin{split}
    m_{N^\prime}^2 &= (2.1 \pm 0.2)~\rm{GeV^2}~, \\
    m_{\Lambda^\prime}^2 &= (2.7 \pm 0.1)~\rm{GeV^2}~.
  \end{split}
\end{equation}

Performing similar analysis for the mass of the $\Sigma$ and $\Xi$ baryons we obtained

\begin{equation}
  \label{eq:massprimesigma}
  \begin{split}
    m_{\Sigma^\prime}^2 &= (2.8 \pm 0.1)~\rm{GeV^2}~, \\
    m_{\Xi^\prime}^2 &= (3.4 \pm 0.2)~\rm{GeV^2}~.
  \end{split}
\end{equation}

Finally, we can determine the residues of the radial excitation baryons. For this aim, we used eq.~\eqref{eq:primesquare}. Using the working region for $M^2$ and at given values of $s_0$ we studied the dependency of the residue square on $\cos{\theta}$ and we obtained the following values:
\begin{equation}
  \label{eq:lambdasquare}
  \begin{split}
    \lambda_{N^\prime}^2 &= (2.0 \pm 0.5) \times 10^{-3}~, \\
    \lambda_{\Sigma^\prime}^2 &= (5.0 \pm 2.0) \times 10^{-3}~, \\
    \lambda_{\Lambda^\prime}^2 &= (5.0 \pm 2.0) \times 10^{-3}~, \\
    \lambda_{\Xi^\prime}^2 &= (9.0 \pm 4.0) \times 10^{-3}~.
  \end{split}
\end{equation}

Comparing our predictions on the mass of the radial excitation with the octet baryons we see that the results are in good agreement with the experimental data. Remember that experimentally the masses of the radial excitations of octet baryons are: $m_{N^\prime} = (1.43 \pm 0.02)~\rm{GeV}$, $m_{\Lambda^\prime} = (1.63 \pm 0.07)~\rm{GeV}$, $m_{\Sigma^\prime} = (1.66 \pm 0.03)~\rm{GeV}$ and $m_{\Xi^\prime} = (1.950 \pm 0.015)~\rm{GeV}~.$ 

We observed that our predictions on mass for the radial excitations are in good agreement with the experimental data. We also calculate the residues for the radial excitations octet baryons.

Our final remark to this section is as follows. We perform our analysis without taking into account the radiative corrections to the two-point correlation function. The one-loop radiative corrections to the two-point correlation function for nucleon for Ioffe current ($\beta = -1$) is calculated in~\cite{Drukarev:2014mua}. When we take into account these corrections, our results change $(5 - 8)\%$.

\section{\label{sec:conclusion}Conclusion}
In present work, we estimate the masses and residues of the first radial excited states of the octet baryons. The QCD sum rules are modified by including the contribution of the first radial excited states in addition to the ground state. Our predictions on the masses of the first radial excited states baryons are in good agreement with the experimental data. Hence, QCD sum rules work quite well not only for ground state baryons but also for the radial excitations. The obtained results for the residues can be checked by studying the electromagnetic or strong transitions of radial excitation to the ground state baryons.

\bibliography{radial_excitations_of_octet_baryons_in_qcd_sum_rules}{}

\begin{thebibliography}{13}%
\makeatletter
\providecommand \@ifxundefined [1]{%
 \@ifx{#1\undefined}
}%
\providecommand \@ifnum [1]{%
 \ifnum #1\expandafter \@firstoftwo
 \else \expandafter \@secondoftwo
 \fi
}%
\providecommand \@ifx [1]{%
 \ifx #1\expandafter \@firstoftwo
 \else \expandafter \@secondoftwo
 \fi
}%
\providecommand \natexlab [1]{#1}%
\providecommand \enquote  [1]{``#1''}%
\providecommand \bibnamefont  [1]{#1}%
\providecommand \bibfnamefont [1]{#1}%
\providecommand \citenamefont [1]{#1}%
\providecommand \href@noop [0]{\@secondoftwo}%
\providecommand \href [0]{\begingroup \@sanitize@url \@href}%
\providecommand \@href[1]{\@@startlink{#1}\@@href}%
\providecommand \@@href[1]{\endgroup#1\@@endlink}%
\providecommand \@sanitize@url [0]{\catcode `\\12\catcode `\$12\catcode
  `\&12\catcode `\#12\catcode `\^12\catcode `\_12\catcode `\%12\relax}%
\providecommand \@@startlink[1]{}%
\providecommand \@@endlink[0]{}%
\providecommand \url  [0]{\begingroup\@sanitize@url \@url }%
\providecommand \@url [1]{\endgroup\@href {#1}{\urlprefix }}%
\providecommand \urlprefix  [0]{URL }%
\providecommand \Eprint [0]{\href }%
\providecommand \doibase [0]{http://dx.doi.org/}%
\providecommand \selectlanguage [0]{\@gobble}%
\providecommand \bibinfo  [0]{\@secondoftwo}%
\providecommand \bibfield  [0]{\@secondoftwo}%
\providecommand \translation [1]{[#1]}%
\providecommand \BibitemOpen [0]{}%
\providecommand \bibitemStop [0]{}%
\providecommand \bibitemNoStop [0]{.\EOS\space}%
\providecommand \EOS [0]{\spacefactor3000\relax}%
\providecommand \BibitemShut  [1]{\csname bibitem#1\endcsname}%
\let\auto@bib@innerbib\@empty
\bibitem [{\citenamefont {Olive}\ \emph {et~al.}(2014)\citenamefont {Olive}
  \emph {et~al.}}]{pdg}%
  \BibitemOpen
  \bibfield  {author} {\bibinfo {author} {\bibfnamefont {K.~A.}\ \bibnamefont
  {Olive}} \emph {et~al.} (\bibinfo {collaboration} {Particle Data Group}),\
  }\href {\doibase 10.1088/1674-1137/38/9/090001} {\bibfield  {journal}
  {\bibinfo  {journal} {Chin. Phys.}\ }\textbf {\bibinfo {volume} {C38}},\
  \bibinfo {pages} {090001} (\bibinfo {year} {2014})}\BibitemShut {NoStop}%
\bibitem [{\citenamefont {Shifman}\ \emph {et~al.}(1979)\citenamefont
  {Shifman}, \citenamefont {Vainshtein},\ and\ \citenamefont
  {Zakharov}}]{shifman1979qcd}%
  \BibitemOpen
  \bibfield  {author} {\bibinfo {author} {\bibfnamefont {M.~A.}\ \bibnamefont
  {Shifman}}, \bibinfo {author} {\bibfnamefont {A.~I.}\ \bibnamefont
  {Vainshtein}}, \ and\ \bibinfo {author} {\bibfnamefont {V.~I.}\ \bibnamefont
  {Zakharov}},\ }\href@noop {} {\bibfield  {journal} {\bibinfo  {journal}
  {Nucl. Phys.}\ }\textbf {\bibinfo {volume} {B147}},\ \bibinfo {pages} {385}
  (\bibinfo {year} {1979})}\BibitemShut {NoStop}%
\bibitem [{\citenamefont {Ioffe}(1981)}]{Ioffe:1981kw}%
  \BibitemOpen
  \bibfield  {author} {\bibinfo {author} {\bibfnamefont {B.~L.}\ \bibnamefont
  {Ioffe}},\ }\href {\doibase 10.1016/0550-3213(81)90315-1,
  10.1016/0550-3213(81)90259-5} {\bibfield  {journal} {\bibinfo  {journal}
  {Nucl. Phys.}\ }\textbf {\bibinfo {volume} {B188}},\ \bibinfo {pages} {317}
  (\bibinfo {year} {1981})},\ \bibinfo {note} {[Erratum: Nucl.
  Phys.B191,591(1981)]}\BibitemShut {NoStop}%
\bibitem [{\citenamefont {Chung}\ \emph {et~al.}(1982)\citenamefont {Chung},
  \citenamefont {Dosch}, \citenamefont {Kremer},\ and\ \citenamefont
  {Schall}}]{chung1982baryon}%
  \BibitemOpen
  \bibfield  {author} {\bibinfo {author} {\bibfnamefont {Y.}~\bibnamefont
  {Chung}}, \bibinfo {author} {\bibfnamefont {H.}~\bibnamefont {Dosch}},
  \bibinfo {author} {\bibfnamefont {M.}~\bibnamefont {Kremer}}, \ and\ \bibinfo
  {author} {\bibfnamefont {D.}~\bibnamefont {Schall}},\ }\href@noop {}
  {\bibfield  {journal} {\bibinfo  {journal} {Nuclear Physics B}\ }\textbf
  {\bibinfo {volume} {197}},\ \bibinfo {pages} {55} (\bibinfo {year}
  {1982})}\BibitemShut {NoStop}%
\bibitem [{\citenamefont {Krasnikov}\ and\ \citenamefont
  {Pivovarov}(1982)}]{krasnikov1982influence}%
  \BibitemOpen
  \bibfield  {author} {\bibinfo {author} {\bibfnamefont {N.}~\bibnamefont
  {Krasnikov}}\ and\ \bibinfo {author} {\bibfnamefont {A.}~\bibnamefont
  {Pivovarov}},\ }\href@noop {} {\bibfield  {journal} {\bibinfo  {journal}
  {Phys. Lett.}\ }\textbf {\bibinfo {volume} {B116}},\ \bibinfo {pages} {168}
  (\bibinfo {year} {1982})}\BibitemShut {NoStop}%
\bibitem [{\citenamefont {Gorishny}\ \emph {et~al.}(1984)\citenamefont
  {Gorishny}, \citenamefont {Kataev},\ and\ \citenamefont
  {Larin}}]{gorishny1984next}%
  \BibitemOpen
  \bibfield  {author} {\bibinfo {author} {\bibfnamefont {S.}~\bibnamefont
  {Gorishny}}, \bibinfo {author} {\bibfnamefont {A.}~\bibnamefont {Kataev}}, \
  and\ \bibinfo {author} {\bibfnamefont {S.}~\bibnamefont {Larin}},\
  }\href@noop {} {\bibfield  {journal} {\bibinfo  {journal} {Phys. Lett.}\
  }\textbf {\bibinfo {volume} {B135}},\ \bibinfo {pages} {457} (\bibinfo {year}
  {1984})}\BibitemShut {NoStop}%
\bibitem [{\citenamefont {Gelhausen}\ \emph {et~al.}(2014)\citenamefont
  {Gelhausen}, \citenamefont {Khodjamirian}, \citenamefont {Pivovarov},\ and\
  \citenamefont {Rosenthal}}]{gelhausen2014radial}%
  \BibitemOpen
  \bibfield  {author} {\bibinfo {author} {\bibfnamefont {P.}~\bibnamefont
  {Gelhausen}}, \bibinfo {author} {\bibfnamefont {A.}~\bibnamefont
  {Khodjamirian}}, \bibinfo {author} {\bibfnamefont {A.}~\bibnamefont
  {Pivovarov}}, \ and\ \bibinfo {author} {\bibfnamefont {D.}~\bibnamefont
  {Rosenthal}},\ }\href@noop {} {\bibfield  {journal} {\bibinfo  {journal} {The
  Eur. Phys. J. C}\ }\textbf {\bibinfo {volume} {74}},\ \bibinfo {pages} {1}
  (\bibinfo {year} {2014})}\BibitemShut {NoStop}%
\bibitem [{\citenamefont {Jiang}\ and\ \citenamefont
  {Zhu}(2015)}]{jiang2015radial}%
  \BibitemOpen
  \bibfield  {author} {\bibinfo {author} {\bibfnamefont {J.-F.}\ \bibnamefont
  {Jiang}}\ and\ \bibinfo {author} {\bibfnamefont {S.-L.}\ \bibnamefont
  {Zhu}},\ }\href@noop {} {\bibfield  {journal} {\bibinfo  {journal} {Phys.
  Rev. D}\ }\textbf {\bibinfo {volume} {92}},\ \bibinfo {pages} {074002}
  (\bibinfo {year} {2015})}\BibitemShut {NoStop}%
\bibitem [{\citenamefont {Lee}\ and\ \citenamefont
  {Liu}(2002)}]{lee2002predictive}%
  \BibitemOpen
  \bibfield  {author} {\bibinfo {author} {\bibfnamefont {F.~X.}\ \bibnamefont
  {Lee}}\ and\ \bibinfo {author} {\bibfnamefont {X.}~\bibnamefont {Liu}},\
  }\href@noop {} {\bibfield  {journal} {\bibinfo  {journal} {Phys. Rev. D}\
  }\textbf {\bibinfo {volume} {66}},\ \bibinfo {pages} {014014} (\bibinfo
  {year} {2002})}\BibitemShut {NoStop}%
\bibitem [{\citenamefont {Aliev}\ \emph {et~al.}(2002)\citenamefont {Aliev},
  \citenamefont {{\"O}zpineci},\ and\ \citenamefont
  {Savc{\i}}}]{aliev2002octet}%
  \BibitemOpen
  \bibfield  {author} {\bibinfo {author} {\bibfnamefont {T.}~\bibnamefont
  {Aliev}}, \bibinfo {author} {\bibfnamefont {A.}~\bibnamefont {{\"O}zpineci}},
  \ and\ \bibinfo {author} {\bibfnamefont {M.}~\bibnamefont {Savc{\i}}},\
  }\href@noop {} {\bibfield  {journal} {\bibinfo  {journal} {Phys. Rev. D}\
  }\textbf {\bibinfo {volume} {66}},\ \bibinfo {pages} {016002} (\bibinfo
  {year} {2002})}\BibitemShut {NoStop}%
\bibitem [{\citenamefont {Narison}(2005)}]{Narison:2005zg}%
  \BibitemOpen
  \bibfield  {author} {\bibinfo {author} {\bibfnamefont {S.}~\bibnamefont
  {Narison}},\ }\href {\doibase 10.1016/j.physletb.2005.08.085} {\bibfield
  {journal} {\bibinfo  {journal} {Phys. Lett.}\ }\textbf {\bibinfo {volume}
  {B626}},\ \bibinfo {pages} {101} (\bibinfo {year} {2005})},\ \Eprint
  {http://arxiv.org/abs/hep-ph/0501208} {arXiv:hep-ph/0501208 [hep-ph]}
  \BibitemShut {NoStop}%
\bibitem [{\citenamefont {Narison}(2009)}]{Narison:2009vy}%
  \BibitemOpen
  \bibfield  {author} {\bibinfo {author} {\bibfnamefont {S.}~\bibnamefont
  {Narison}},\ }\href {\doibase 10.1016/j.physletb.2009.01.062} {\bibfield
  {journal} {\bibinfo  {journal} {Phys. Lett.}\ }\textbf {\bibinfo {volume}
  {B673}},\ \bibinfo {pages} {30} (\bibinfo {year} {2009})},\ \Eprint
  {http://arxiv.org/abs/0901.3823} {arXiv:0901.3823 [hep-ph]} \BibitemShut
  {NoStop}%
\bibitem [{\citenamefont {Drukarev}\ \emph {et~al.}(2015)\citenamefont
  {Drukarev}, \citenamefont {Ryskin},\ and\ \citenamefont
  {Sadovnikova}}]{Drukarev:2014mua}%
  \BibitemOpen
  \bibfield  {author} {\bibinfo {author} {\bibfnamefont {E.~G.}\ \bibnamefont
  {Drukarev}}, \bibinfo {author} {\bibfnamefont {M.~G.}\ \bibnamefont
  {Ryskin}}, \ and\ \bibinfo {author} {\bibfnamefont {V.~A.}\ \bibnamefont
  {Sadovnikova}},\ }\href {\doibase 10.1134/S1063778815070054} {\bibfield
  {journal} {\bibinfo  {journal} {Phys. Atom. Nucl.}\ }\textbf {\bibinfo
  {volume} {78}},\ \bibinfo {pages} {849} (\bibinfo {year} {2015})},\ \bibinfo
  {note} {[Yad. Fiz.78,no.10,903(2015)]},\ \Eprint
  {http://arxiv.org/abs/1410.6282} {arXiv:1410.6282 [hep-ph]} \BibitemShut
  {NoStop}%
\end{thebibliography}%
\newpage

\end{document}